\documentclass[conference]{IEEEtran}
\IEEEoverridecommandlockouts

\usepackage{cite}
\usepackage{amsmath,amssymb,amsfonts}
\usepackage{algorithmic}
\usepackage{graphicx}
\usepackage{textcomp}
\usepackage{xcolor}
\usepackage{array}
\usepackage{url}

\def\BibTeX{{\rm B\kern-.05em{\sc i\kern-.025em b}\kern-.08em
    T\kern-.1667em\lower.7ex\hbox{E}\kern-.125emX}}

\begin{document}

\title{OmniSynth: A Meta-Evaluation Framework for Benchmarking the Quality of Missingness-Inclusive Synthetic Health Data}

\author{}

\author{\IEEEauthorblockN{1\textsuperscript{st} Annabel Zhu}
\IEEEauthorblockA{\textit{McGill University} \\
Montreal, Canada \\
annabel.zhu@mail.mcgill.ca}
\and
\IEEEauthorblockN{2\textsuperscript{nd} Océane Doremus}
\IEEEauthorblockA{\textit{University of Bordeaux, INSERM U1219} \\
Bordeaux, France \\
oceane.doremus@u-bordeaux.fr}
\and
\IEEEauthorblockN{3\textsuperscript{rd} Ariel Guerra-Adames}
\IEEEauthorblockA{\textit{University of Bordeaux, INSERM U1219} \\
Bordeaux, France \\
ariel.guerra-adames@u-bordeaux.fr}
\and
\IEEEauthorblockN{4\textsuperscript{th} Marta Avalos Fernandez}
\IEEEauthorblockA{\textit{University of Bordeaux, INSERM U1219} \\
Bordeaux, France \\
marta.avalos-fernandez@u-bordeaux.fr}
\and
\IEEEauthorblockN{5\textsuperscript{th} Emmanuel Lagarde}
\IEEEauthorblockA{\textit{University of Bordeaux, INSERM U1219} \\
Bordeaux, France \\
emmanuel.lagarde@u-bordeaux.fr}
\and
\IEEEauthorblockN{6\textsuperscript{th} Cédric Gil-Jardiné}
\IEEEauthorblockA{\textit{University of Bordeaux, INSERM U1219} \\
Bordeaux, France \\
cedric.gil-jardine@chu-bordeaux.fr}
\and
\IEEEauthorblockN{7\textsuperscript{th} Vianney Jouhet}
\IEEEauthorblockA{\textit{Bordeaux University Hospital, INSERM U1219} \\
Bordeaux, France \\
vianney.jouhet@chu-bordeaux.fr}
}
\maketitle

\begin{abstract}
Synthetic data are an emerging solution to the reproducibility crisis in machine learning for health, driven by data privacy restrictions surrounding electronic health records. Yet no consensus exists on how to evaluate synthetic health data quality, existing metrics functionally overlook missingness despite its prevalence in clinical data, and benchmarking frameworks have never been empirically validated. We introduce OmniSynth, a modular Python benchmarking suite that evaluates synthetic tabular data across fidelity, missingness similarity, utility, and privacy---the first framework to explicitly evaluate missingness as a distinct dimension of synthetic data fidelity. OmniSynth further proposes a meta-evaluation procedure that stress-tests evaluation performance under controlled degradation scenarios, providing a systematic characterization of how an evaluation framework behaves across datasets, sample sizes, and variable types. Applying this procedure to OmniSynth in two health datasets (Pima Indians Diabetes; MIMIC-IV-ED), we found that scores responded monotonically to degradation, three of four novel missingness metrics captured induced missing data patterns, and individual metric contributions to the scoring system were empirically justified.
\end{abstract}

\begin{IEEEkeywords}
synthetic data, evaluation framework, missingness, health data, meta-evaluation, benchmarking
\end{IEEEkeywords}

\section{Introduction}

Access to large-scale health data remain heavily restricted due to privacy regulations and the sensitive nature of electronic health records (EHRs), limiting reproducibility and secondary use of medical machine learning research \cite{mcdermottReproducibilityMachineLearning2021,semmelrockReproducibilityMachinelearningbasedResearch2025}. Synthetic data generation has therefore emerged as a promising approach for enabling data sharing while reducing disclosure risks \cite{gonzalesSyntheticDataHealth2023}. In public health and clinical research, high-quality synthetic data could facilitate reproducible analyses, external validation, and collaborative model development without exposing identifiable patient records \cite{elemamUtilityMetricsEvaluating2022,gonzalesSyntheticDataHealth2023}.

Despite the rapid development of synthetic data generation methods, there is currently no consensus on how synthetic health data should be evaluated \cite{beduschiSyntheticDataProtection2024,hernandezComprehensiveEvaluationFramework2025,yanMultifacetedBenchmarkingSynthetic2022}. The quality of synthetic medical data generally revolves around three main axes: fidelity, which refers to the closeness of the generated data to the original data; utility, which assesses the relevance of synthetic data for specific applications; and privacy, which ensures that the synthetic data cannot be used to reconstruct or reidentify individuals in the original dataset \cite{santangeloHowGoodYour2025}. In the absence of evaluation standards for synthetic data quality, existing frameworks typically use a broad range of metrics to comprehensively measure various aspects of synthetic data against the original real data. These can be used to rank synthetic datasets or synthetic data generation methods against one another for a specified use-case \cite{lautrupSynthEvalFrameworkDetailed2025,santangeloHowGoodYour2025,yanMultifacetedBenchmarkingSynthetic2022}. However, most proposed evaluation and benchmarking frameworks in the literature are tested directly on synthetic data generation methods as a proof of concept, while the behaviour of the frameworks under controlled conditions remains untested. In this work, we refer to this process as the \textit{meta-evaluation}: the systematic assessment of evaluation behaviour under predefined degradation scenarios designed to stress-test specific quality dimensions.

Another important limitation of current evaluation frameworks is the treatment of missing data. Missingness is ubiquitous in real-world clinical and public health datasets and may itself contain clinically informative signals when missingness mechanisms are not completely random \cite{beaulieu-jonesCharacterizingManagingMissing2018,wangPreservingMissingData2023,santosGeneratingSyntheticMissing2019}. Previous work has also shown that accounting for missing data, rather than imputing or relying on complete case analysis, may improve the quality of machine learning models \cite{wangPreservingMissingData2023}. Nevertheless, most existing synthetic data metrics either ignore missing values or rely on complete-case analysis prior to evaluation. We therefore consider missingness similarity as an explicit dimension of synthetic health data quality.

To address these gaps, we propose a meta-evaluation methodology for systematically assessing synthetic data evaluation frameworks under controlled degradation scenarios. We instantiate this methodology in OmniSynth, a modular Python framework for evaluating synthetic tabular health data across fidelity, utility, privacy, and missingness dimensions. OmniSynth additionally introduces four novel metrics for comparing missingness patterns between real and synthetic datasets.

In summary, this work makes the following contributions:
\begin{itemize}
    \item We propose a meta-evaluation methodology for assessing the sensitivity, stability, and reliability of synthetic data evaluation frameworks under controlled degradation scenarios.
    \item We introduce OmniSynth, a modular framework for evaluating synthetic tabular health data across fidelity, utility, privacy, and missingness similarity dimensions.
    \item We define, implement, and evaluate four novel metrics for comparing missingness patterns between real and synthetic datasets.
\end{itemize}

\section{Related Works}

\subsection{Evaluation frameworks and metrics} 

Existing synthetic data evaluation frameworks implement a broad range of metrics spanning fidelity, utility, and privacy dimensions (Table~\ref{tab:frameworks}). Fidelity metrics generally assess statistical similarity between real and synthetic data distributions through marginal distributions, pairwise dependencies, or multivariable distinguishability measures; utility metrics evaluate downstream analytical performance; and privacy metrics estimate disclosure and re-identification risks.

Current frameworks primarily focus on metric implementation and benchmarking workflows for comparing synthetic datasets or generative models, but generally do not assess the behaviour of the benchmark itself. Consequently, the behaviour of evaluation metrics and aggregated benchmark scores under controlled degradations remains poorly characterized, making score interpretation, calibration, and comparison difficult across datasets and experimental settings.

\begin{table}[htbp]
\caption{Comparison of Existing Evaluation Tools for Synthetic Data (SD).}
\label{tab:frameworks}
\begin{center}
\begin{tabular}{|p{1.4cm}|p{2.7cm}|p{1.7cm}|p{1.3cm}|}
\hline
\textbf{Tool} & \textbf{Description} & \textbf{Axes} & \textbf{Meta-eval.} \\
\hline
SDMetrics \cite{sdmetrics} & Evaluation metric suite. & fidelity, privacy & no \\
\hline
Synthcity \cite{qian2023synthcity} & SD generation and evaluation metric suite. & fidelity, utility, privacy & no \\
\hline
SynthEval \cite{lautrupSynthEvalFrameworkDetailed2025} & Evaluation metric suite and benchmarking framework. & fidelity, utility, privacy & no \\
\hline
SynthRO \cite{santangeloHowGoodYour2025} & Evaluation, benchmarking framework with visualization dashboard. & fidelity, utility, privacy & no \\
\hline
\textbf{OmniSynth} & Evaluation and benchmarking framework with stress-testing methodology. & fidelity, utility, privacy, missingness & yes \\
\hline
\end{tabular}
\end{center}
\end{table}

\subsection{Missingness-aware evaluation} 

Although missingness mechanisms such as missing completely at random (MCAR), missing at random (MAR), and missing not at random (MNAR) are pervasive in clinical and public health datasets \cite{rubinInferenceMissingData1976}, existing synthetic data evaluation frameworks rarely assess whether missingness patterns are preserved. Current approaches are typically limited to descriptive statistics such as per-variable missingness rates \cite{sdmetrics}.

Taken together, these limitations motivate the need for systematic benchmark validation methodologies capable of assessing metric sensitivity, calibration, robustness, and missingness preservation under controlled degradation scenarios.

\section{OmniSynth: Evaluation and Benchmarking Framework}

\subsection{OmniSynth workflow and scoring}

OmniSynth is a modular Python framework for evaluating synthetic tabular health data across fidelity, utility, privacy, and missingness similarity dimensions. Given a real dataset $X$ and a synthetic dataset $Y$, the framework computes a set of normalized metric scores grouped by evaluation axis. Metric scores are aggregated into axis-specific scores using weighted averages, which are then combined into a global composite score across evaluation axes (Fig.~\ref{fig:flowchart}). By default, all evaluation axes receive equal weights.

All metric scores are normalized to the range $[0,1]$, where 0 represents the worst possible score and 1 the best. The resulting scores are used to characterize synthetic data quality and support comparative evaluation across datasets or generation settings. The framework and meta-evaluation pipeline are publicly available at: \url{https://github.com/annabelz/OmniSynth}

\begin{figure*}[htbp]
\centerline{\includegraphics[width=0.80\textwidth]{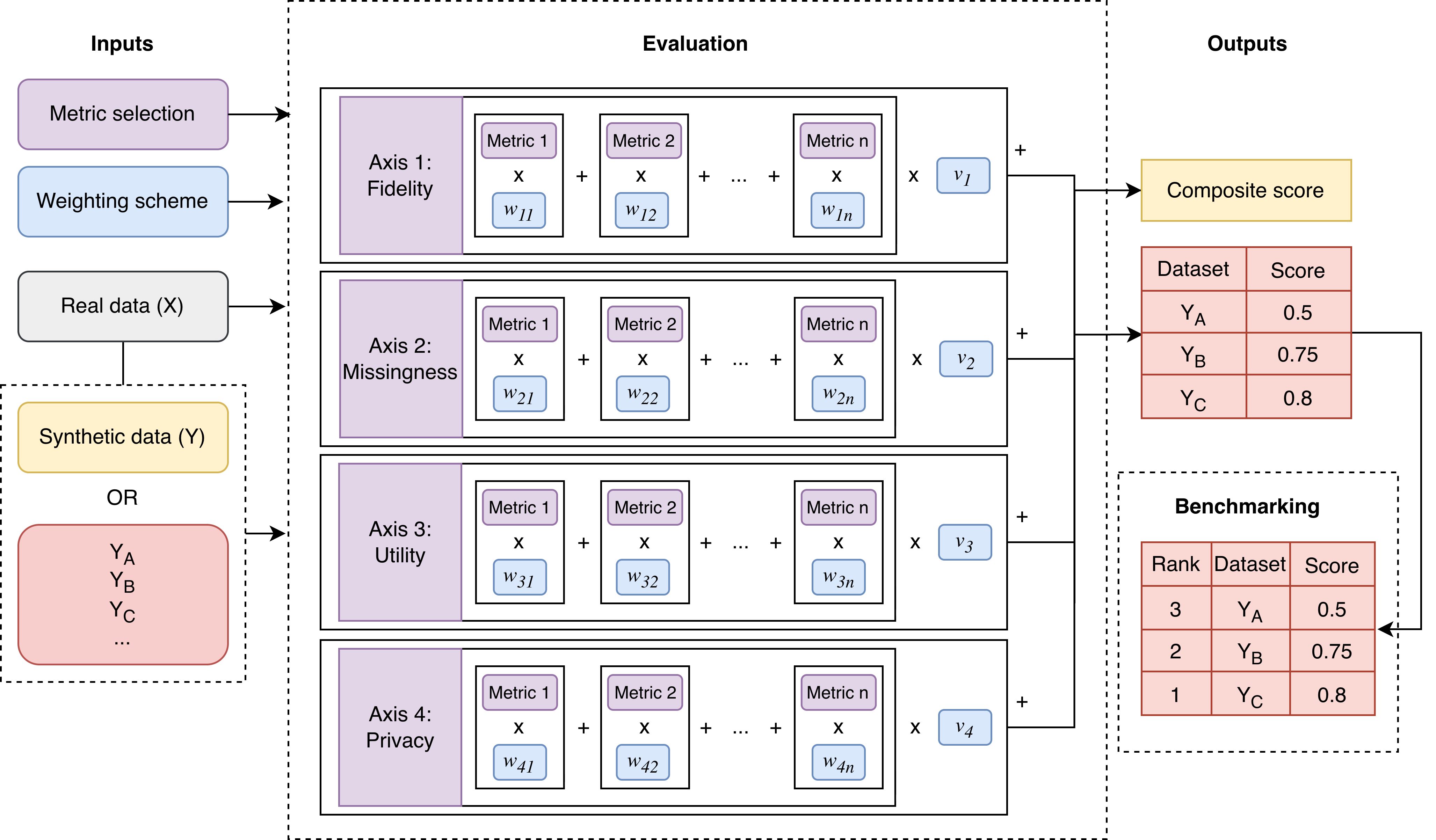}}
\caption{OmniSynth flowchart where $w$ denotes the weights for individual metrics used in the axis summary score, and $v$ denotes the weights for evaluation axes used in the overall composite score for benchmarking.}
\label{fig:flowchart}
\end{figure*}

\subsection{Selection of evaluation metrics}

The selected metrics cover fidelity, utility, privacy, and missingness dimensions, and were chosen for their applicability to mixed tabular health data containing both numerical and categorical variables (Table~\ref{tab:metrics}). All metrics are normalized to the range $[0,1]$, with higher values indicating better synthetic data quality. OmniSynth additionally introduces four novel metrics designed to evaluate similarity in missingness patterns between real and synthetic datasets.

\begin{table*}[htbp]
\caption{Description of Metrics Implemented in Benchmarking Framework. Computational complexity is given where $n$ = rows, $p$ = columns, $p'$ = columns after one-hot encoding categorical variables; and for classifier-based metrics, $T$ = n\_estimators, $k$ = cv\_folds, $D$ = max\_depth, $I$ = max\_iterations.}
\label{tab:metrics}
\begin{center}
\begin{tabular}{|p{1.3cm}|p{3.5cm}|p{9.8cm}|p{1.6cm}|}
\hline
\textbf{Axis} & \textbf{Metric (\underline{Function})} & \textbf{Implementation description} & \textbf{Complexity} \\
\hline
Fidelity
  & Dimension-wise distribution (\underline{HellingerDistance})
  & Mean complement of Hellinger score across columns, measuring distance between marginal feature distributions in real and synthetic data \cite{hernandezComprehensiveEvaluationFramework2025}.
  & $O(n \cdot p)$ \\
\cline{2-4}
  & Column-wise correlation (\underline{PCD})
  & Mean absolute difference in phi-k correlations between variable pairs, capturing differences in pairwise feature correlations \cite{baakNewCorrelationCoefficient2020,hernandezComprehensiveEvaluationFramework2025}.
  & $O(n \cdot p^2)$ \\
\cline{2-4}
  & Classifier differentiation performance (\underline{AuRoc})
  & $k$-fold cross-validated area under the receiver operating characteristic curve (AUROC) of a random forest classifier trained to distinguish real from synthetic samples; lower discriminability indicates higher fidelity \cite{hernandezComprehensiveEvaluationFramework2025}.
  & $O((1{+}k){\cdot}T{\cdot}$
  $p'{\cdot}n{\cdot}D)$ \\
\cline{2-4}
  & Propensity score MSE (\underline{pMSE})
  & Average propensity score MSE from a logistic regression predicting whether each record belongs to the synthetic dataset \cite{snokeGeneralSpecificUtility2017,wooGlobalMeasuresData2009}.
  & $O(k{\cdot}n{\cdot}p'{\cdot}I)$ \\
\hline
Missingness similarity
  & Missingness rate difference (\underline{MissRateDiff})
  & Mean absolute difference in per-column missingness proportions between real and synthetic data.
  & $O(n \cdot p)$ \\
\cline{2-4}
  & Missingness set distribution (\underline{MissSet})
  & Total variation distance between the distributions over joint missingness patterns in real and synthetic data.
  & $O(n \cdot p)$ \\
\cline{2-4}
  & Missingness dependency structure (\underline{MissDependency})
  & Mean absolute difference in pairwise correlations of joint missingness indicators between real and synthetic data \cite{suzenWhatHidingMedicines2023}.
  & $O(n{\cdot}p'^2{\cdot}I)$ \\
\cline{2-4}
  & Missingness classifier performance (\underline{MissAuRoc})
  & Per-variable logistic regression predicting missingness; AUROC compared between real and synthetic data to assess whether missingness mechanisms are faithfully reproduced \cite{beaulieu-jonesCharacterizingManagingMissing2018}.
  & $O(n \cdot p^2)$ \\
\hline
Privacy
  & Exact match rate (\underline{ExactMatch})
  & Proportion of synthetic records with zero distance to any real record after preprocessing, detecting direct memorization \cite{platzerSyntheticDataAnonymisation2021}.
  & $O(n^2 \cdot p')$ \\
\cline{2-4}
  & Distance to closest record (\underline{DCR})
  & Proportion of synthetic records whose nearest-neighbour distance exceeds a threshold derived from real-to-real distances \cite{platzerSyntheticDataAnonymisation2021}.
  & $O(n^2 \cdot p')$ \\
\cline{2-4}
  & Nearest-neighbour distance ratio (\underline{NNDR})
  & Ratio of distances to the two nearest real neighbours per synthetic record, detecting targeted memorization through distance asymmetry \cite{murtagh2018}.
  & $O(n^2 \cdot p')$ \\
\cline{2-4}
  & Nearest-neighbour adversarial accuracy (\underline{NNAA})
  & Adversarial nearest-neighbour comparison assessing whether real records are distinguishable from synthetic records \cite{yaleAssessingPrivacyMachine2020,giomiCloserLookPrivacy2023}.
  & $O(n^2 \cdot p')$ \\
\cline{2-4}
  & Singling-out risk (\underline{SinglingOutRisk})
  & Proportion of real records located within a small neighbourhood of a synthetic record, measuring re-identification risk \cite{giomiCloserLookPrivacy2023}.
  & $O(n^2 \cdot p')$ \\
\cline{2-4}
  & Membership inference AUROC (\underline{MembershipAuRoc})
  & Cross-validated AUROC of a random forest classifier distinguishing real from synthetic records, measuring vulnerability to membership inference attacks \cite{stadlerSyntheticDataPrivacy2022}.
  & $O((1{+}k){\cdot}T{\cdot}$
  $p'{\cdot}n{\cdot}D)$ \\
\cline{2-4}
  & Attribute disclosure risk (\underline{AttrDisclosure})
  & Decision tree trained on synthetic data to predict sensitive attributes, evaluated on real data \cite{wooGlobalMeasuresData2009,stadlerSyntheticDataPrivacy2022}.
  & $O(k{\cdot}n{\cdot}p'{\cdot}D)$ \\
\hline
Utility
  & TSTR performance gap (\underline{TSTRDelta})
  & AUROC (classification) or $R^2$ (regression) gap between models trained on real (train-real-test-real; TRTR) and synthetic (train-synthetic-test-real; TSTR) data; score of 1 indicates that the synthetic data train an equally performing model \cite{jordonPATEGANGeneratingSynthetic2018}.
  & $O(k{\cdot}n{\cdot}p'{\cdot}I)$ \\
\cline{2-4}
  & Calibration similarity (\underline{CalSimilarity})
  & Brier score compared between TRTR and TSTR models on the same real held-out test set; score of 1 indicates equal calibration (classification only).
  & $O(k \cdot n)$ \\
\cline{2-4}
  & Feature importance alignment (\underline{FISimilarity})
  & Spearman rank correlation between normalised feature importances from models trained on real and synthetic data; score of 1 indicates identical feature rankings.
  & $O(k{\cdot}n{\cdot}p'{\cdot}D)$ \\
\hline
\end{tabular}
\end{center}
\end{table*}

\section{OmniSynth: Meta-Evaluation}
\label{sec:meta-eval}

We designed a novel methodology for assessing the performance and behaviour of the synthetic data evaluation across controlled conditions to stress-test each quality axis (Table~\ref{tab:scenarios}). These scenarios were applied to complete-case datasets at different sample sizes to manually generate multiple test synthetic datasets per scenario. The OmniSynth evaluation was then conducted on each test synthetic dataset, comparing against the original sample from the complete-case data.

The testing scenarios were designed to assess the baseline performance of the evaluation benchmark under controlled settings for each data quality axis (Table~\ref{tab:scenarios}). To assess changes in the fidelity score, the ideal score results from the real and synthetic datasets being exactly the same--this becomes the base scenario. We defined fidelity scenarios to test different aspects of the fidelity scoring as manipulations of the base scenario. Similarly, for the missingness similarity, the ideal score results from the real and synthetic datasets having the same missingness pattern, and missingness scenarios were designed to test changes in missingness similarity scoring. For privacy evaluation, scenarios were constructed to represent increasing levels of record-level memorization and linkability to real individuals, ranging from exact dataset replication to progressively transformed records through column shuffling, feature scaling, and attribute swapping.

\subsection{Datasets}

The meta-evaluation was conducted on two tabular health datasets: the Pima Indians Diabetes dataset (PID) \cite{diabetes_34,smith1988pima} and MIMIC-IV-ED \cite{PhysioNet-mimic-iv-ed-2.2}. PID is a small open-access dataset containing only numerical variables and complete records, making it suitable as a proof-of-concept setting. MIMIC-IV-ED contains emergency department admissions with mixed numerical and categorical variables as well as missing values. Data from the \texttt{edstays}, \texttt{triage}, and \texttt{diagnosis} tables were merged using \texttt{subject\_id} and \texttt{stay\_id}, which were subsequently removed from the final dataset.

\subsection{Experimental setup}
\label{sec:exp-setup}
The controlled degradation scenarios were designed to evaluate several expected properties of a robust synthetic data benchmark. First, benchmark scores were expected to exhibit monotonic behaviour, with progressively stronger perturbations producing lower scores. Second, metrics were expected to demonstrate sensitivity to targeted perturbations affecting the corresponding quality axis. Third, benchmark scores were expected to remain relatively stable across replicates generated under identical conditions. Finally, the baseline scenario--corresponding to the absence of perturbation--was expected to produce near-optimal scores across fidelity and missingness similarity evaluation axes.

Each testing scenario was generated 100 times for PID, across random samples of size $n = 100$, $n = 500$, and $n = 768$ (full sample). Fidelity scenarios 2 and 4 were not generated for PID because it did not contain categorical variables. For MIMIC-IV-ED, samples were generated 100 times for each scenario at sizes $n = 500$, $n = 1\,000$, and $n = 10\,000$.

The configuration files for the meta-evaluation across both datasets were constructed as described, with a \texttt{random\_seed} of 42 and all weights set to their default values. The meta-evaluation command was then run using a virtual environment: \texttt{OmniSynth meta-eval --config configs/meta\_eval\_config.yaml}.

Experiments were run on a personal machine equipped with an Intel Xeon Gold 5220R CPU @ 2.20\,GHz, 132\,GB RAM, and NVIDIA-SMI 535.288.01.

\begin{table}[htbp]
\caption{Description of Scenarios for Meta-Evaluation.}
\label{tab:scenarios}
\begin{center}
\begin{tabular}{|p{1.3cm}|p{2.2cm}|p{3.8cm}|}
\hline
\textbf{Axis} & \textbf{Scenario} & \textbf{Description} \\
\hline
Baseline & Ideal fidelity and missingness similarity scenario. Worst case for privacy. & No change from input dataset. \\
\hline
Fidelity
  & F1: Low noise & Inject low random noise to all variables (numerical and categorical). \\
\cline{2-3}
  & F2: Moderate noise & Inject moderate random noise to all variables (numerical and categorical). \\
\cline{2-3}
  & F3: Low numerical noise & Inject low random noise to numerical variables only, preserving categorical correlations. \\
\cline{2-3}
  & F4: Moderate numerical noise & Inject moderate noise to numerical variables only, preserving categorical correlations. \\
\cline{2-3}
  & F5: Bivariable correlation noise & Induce a correlation between a pair of variables. \\
\hline
Missingness similarity
  & M1: Low MCAR & Inject 10\% missingness at random throughout the input dataset. \\
\cline{2-3}
  & M2: Moderate MCAR & Inject 20\% missingness at random throughout the input dataset. \\
\cline{2-3}
  & M3: High MCAR & Inject 30\% missingness at random throughout the input dataset. \\
\cline{2-3}
  & M4: MAR & Induce a correlation between missingness in one variable, based on observed values of another. \\
\cline{2-3}
  & M5: MNAR & Inject random missingness in a variable for a specified range of values. \\
\hline
Privacy
  & P1: Column shuffle & Randomly permute each column independently. \\
\cline{2-3}
  & P2: Discrete perturbation & Inject small perturbations to numerical variables while preserving categorical variables. \\
\cline{2-3}
  & P3: Feature-wise scaling & Apply random scaling factors to numerical and ordinal variables. \\
\cline{2-3}
  & P4: Attribute swapping & Randomly swap values between records for a subset of numerical and ordinal variables. \\
\hline
\end{tabular}
\end{center}
\end{table}

\section{Results}

\subsection{Evaluation response to controlled degradation}

We first assessed whether benchmark scores exhibited monotonic degradation across progressively perturbed fidelity and missingness scenarios. Fig.~\ref{fig:sample_size_score} summarizes fidelity, missingness, and privacy benchmark scores across datasets and sample sizes. Specifically, we compared fidelity scenarios 1 through 4 to the baseline, and missingness scenarios 1 through 3. As expected, the fidelity scores decreased from Baseline to F1 to F2, then from Baseline to F3 to F4 in all sample sizes and datasets. However, privacy scores decreased from F1 to F2 and from F3 to F4.

In missingness similarity scenarios, we also observed an incremental decrease in missingness scoring from baseline through M1 to M3 in all sample sizes for both datasets. Conversely, privacy scores increased as more random missingness was introduced in these scenarios, though this effect was not consistent for MIMIC-IV-ED sample sizes.


\subsection{Evaluation score stability across sample sizes}

\begin{figure*}[htbp]
\centerline{\includegraphics[width=.95\textwidth]{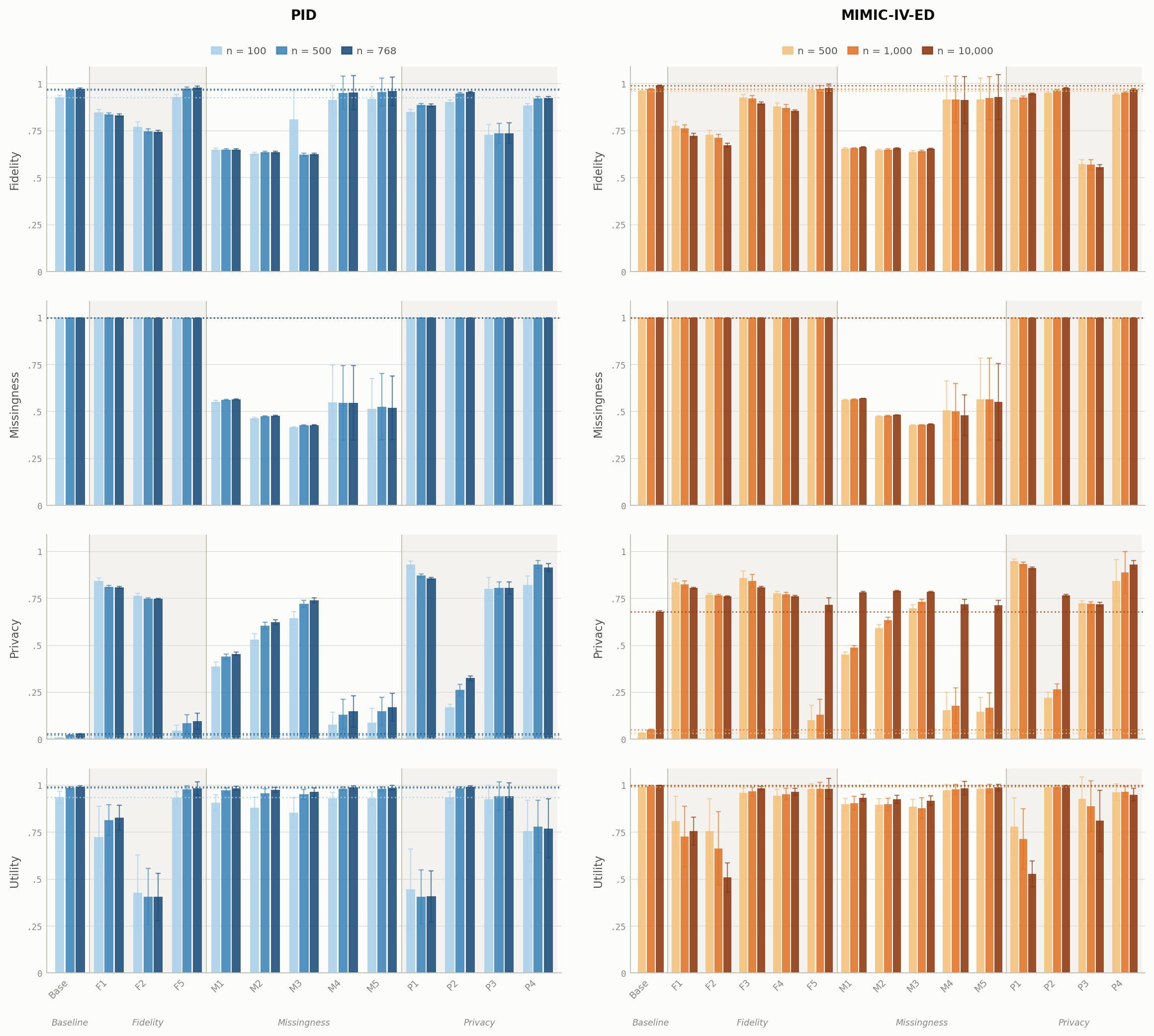}}
\caption{Mean scores within the same dataset, across sample size and scenarios. Baseline scores are shown across the plots in dotted lines. Error bars represent ±1 standard deviation; extensions beyond the theoretical range [0, 1] occur because standard deviations are displayed symmetrically around the mean, although the underlying scale is bounded.}
\label{fig:sample_size_score}
\end{figure*}

\begin{figure*}[htbp]
\centerline{\includegraphics[width=\textwidth]{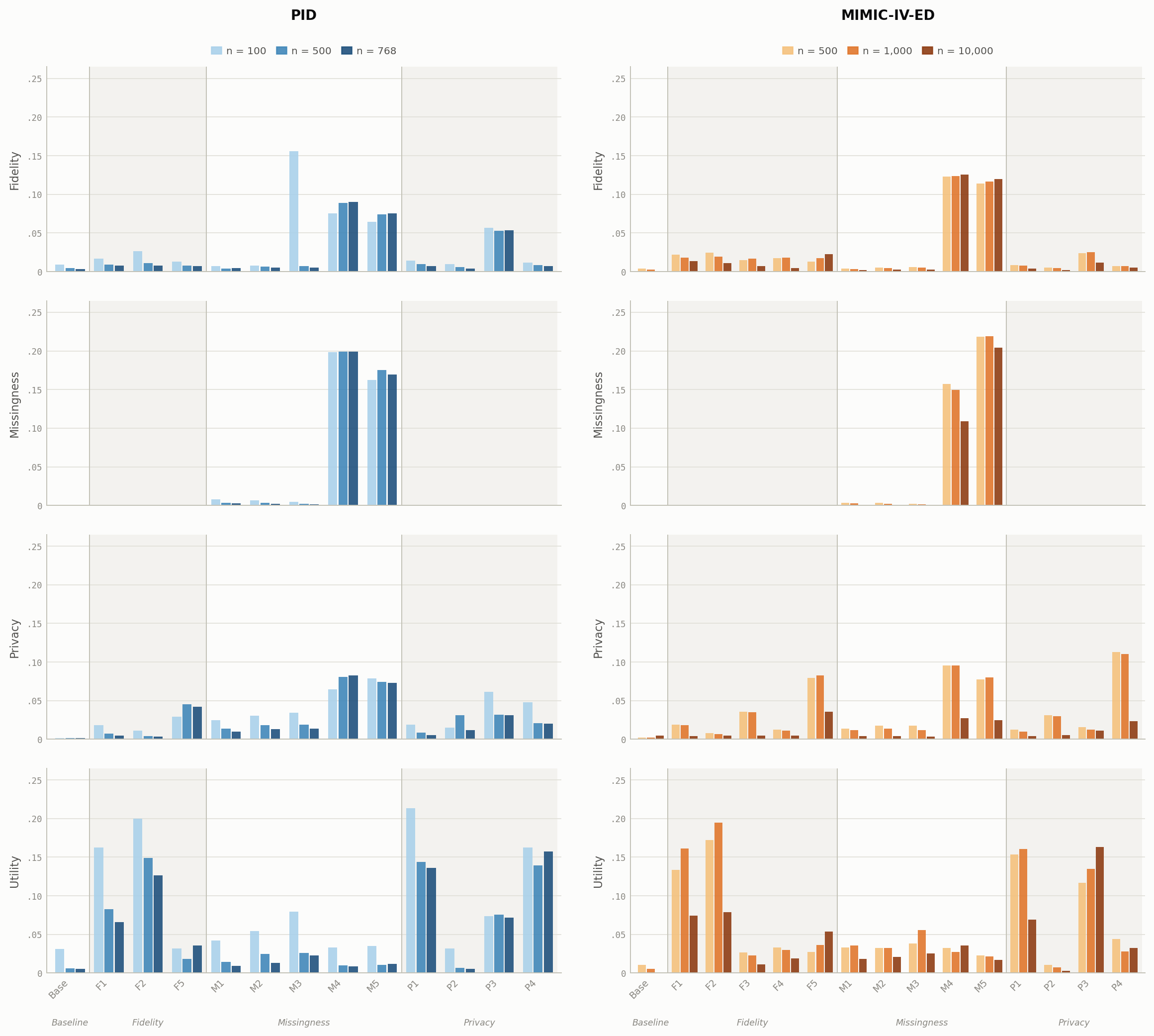}}
\caption{Standard deviation of scores across scenarios for PID and MIMIC-IV-ED datasets, by sample size. PID and MIMIC-IV-ED scenarios have 100 replicates each.}
\label{fig:std}
\end{figure*}

We evaluated benchmark score stability across sample sizes and replicates. In Fig.~\ref{fig:sample_size_score}, we found that the fidelity and missingness scores were relatively stable within each scenario for both datasets $n = 100$. The privacy score was more variable across sample sizes, particularly for MIMIC-IV-ED's $n = 10\,000$ samples (see Fig.~\ref{fig:std})

We expected larger samples to yield smaller standard deviations ($n = 768$ in PID; $n = 10\,000$ in MIMIC-IV-ED), and this trend generally held, as seen in Fig.~\ref{fig:std}. However, large standard deviations persisted across privacy, fidelity, and utility scores, and missingness scenarios, warranting further investigation.

\subsection{Metric contribution to scoring}

\begin{figure*}[htbp]
\centerline{\includegraphics[width=\textwidth]{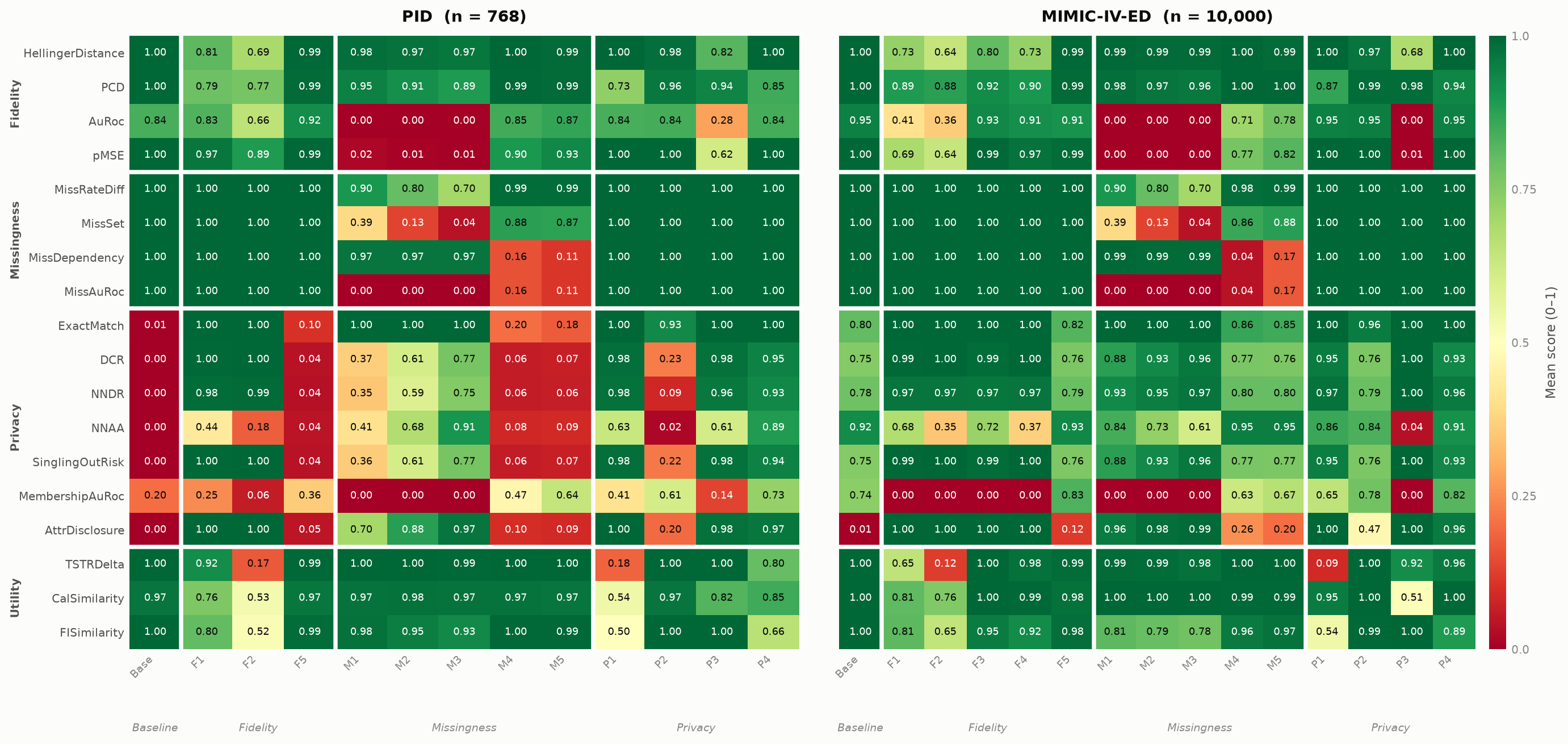}}
\caption{Heatmap of average metric results for the replicates of PID at $n = 768$ and MIMIC-IV-ED at $n = 10\,000$ across all scenarios.}
\label{fig:metric_heatmap}
\end{figure*}

We investigated the contribution of each metric to the scores in each scenario to ensure that metrics responded appropriately to the scenario designs. In Fig.~\ref{fig:metric_heatmap}, we observed that PCD scores for both datasets did not decrease noticeably in scenario F5, which was designed to detect changes in bivariable correlations. We also found that the novel metrics for missingness, MissSet and MissDependency, correctly responded to the progressive MCAR in scenarios M1-M3. MissDependency was relatively unaffected in both datasets when missingness was completely at random, but showed deficiencies in detecting dependencies that were introduced in M4 and M5. The current implementation for MissAuRoc scores 0 across all missingness scenarios since each scenario evaluates a dataset with missingness against a dataset without missingness, though some noise appears in M4 and M5 of PID. Further experiments with datasets containing missingness are required to gauge the performance of these metrics.

The multivariate fidelity metrics (AuRoc, pMSE) appropriately responded to degradations in fidelity scenarios, with stronger responses to MIMIC-IV-ED potentially due to the treatment of categorical variables. Moreover, these returned low values for degraded missingness similarity in scenarios M1-M3. These metrics rely on the performance of the chosen classifier, which may cause variability in scoring.

Most privacy metrics, except for NNAA and MembershipAuRoc, had no response to incremental noise being introduced in the fidelity scenarios F1-F2 and F3-F4. However, all privacy metrics responded to dependency changes in fidelity and missing values (scenarios F5, M4, M5), with smaller responses in the MIMIC-IV-ED samples. In the privacy scenarios, responses were generally consistent across datasets, though noticeably more sensitive in PID.

All utility metrics were sensitive to all-variable fidelity noise in F1-F2, were generally consistent for missingness scenarios (except for FISimilarity in MIMIC-IV-ED), and had varied responses to privacy scenarios. This may be indicative of the three metrics capturing different aspects of data utility, though further investigations and controlled utility-focused testing are needed.
 
\section{Discussion}

Our results showed that OmniSynth generally exhibited the expected behaviour under controlled degradation scenarios while also exposing important limitations in several widely used synthetic data evaluation metrics. Across fidelity and missingness perturbations, benchmark scores demonstrated monotonic degradation and reasonable stability across replicates, supporting the validity of the proposed meta-evaluation methodology. At the same time, some multivariate classifier-based metrics remained unstable, suggesting that benchmark scores may partly reflect properties of the evaluation metrics themselves rather than  differences in synthetic data quality.

Metric behaviour also appeared sensitive to dataset composition and mixed categorical-numerical feature spaces, highlighting the importance of evaluating metric calibration and robustness before interpreting aggregate benchmark scores. Our findings suggest a distinction between two kinds of cross-dataset comparison. Relative patterns--such as which metrics were more or less sensitive to a given perturbation, or whether scores degraded monotonically--were broadly consistent between PID and MIMIC-IV-ED and can reasonably inform generalizable claims about metric behaviour. Absolute score magnitudes, by contrast, were not directly comparable across datasets: the same nominal perturbation (e.g., 10\% MCAR) produced different score magnitudes in PID versus MIMIC-IV-ED, likely reflecting differences in dimensionality, variable type composition, and baseline correlational structure between the two datasets. We therefore recommend that benchmark scores be treated as measures of relative quality within a given dataset and generation setting, rather than as absolute, cross-dataset benchmarks of synthetic data quality.

More broadly, this work highlights benchmark validation itself as an important but largely overlooked problem in synthetic health data evaluation. Future work will apply the proposed meta-evaluation methodology to existing frameworks, generate additional composite and utility-oriented degradation scenarios, and conduct weight sensitivity analyses with varying composite score weighting. Finally, improving synthetic health data evaluation may contribute to more reproducible and collaborative machine learning research in healthcare.

\section{Limitations}

This work has several limitations. First, classifier-based metrics introduce substantial computational overhead for large and heterogeneous datasets, which considerably increased computation times for MIMIC-IV-ED evaluations and limited the sample sizes used in the experiments. Second, some metrics, particularly MissAuRoc and multivariate classifier-based fidelity metrics, exhibited instability or limited interpretability under certain stress-testing scenarios. Third, although the proposed meta-evaluation covers fidelity, privacy, and missingness perturbations, additional composite degradation scenarios and utility-oriented stress tests remain to be explored, as well as incrementally degraded privacy scenarios. In particular, the meta-evaluation of downstream utility metrics was not extensively investigated in this work. Finally, OmniSynth does not yet include fairness-aware evaluation metrics, despite the risk that synthetic data may reproduce or amplify biases present in the original datasets.

\section{Concluding Remarks}

This work introduced a meta-evaluation methodology for systematically stress-testing synthetic health data evaluation frameworks under controlled degradation scenarios. Through experiments on two healthcare datasets, we showed that benchmark behaviour itself can reveal important limitations in metric sensitivity, calibration, and stability, particularly for multivariate classifier-based metrics. We additionally proposed four novel missingness-aware metrics and integrated them into OmniSynth, a modular framework for evaluating synthetic tabular health data across fidelity, utility, privacy, and missingness dimensions. Overall, our findings highlight the importance of validating evaluation frameworks themselves before relying on benchmark scores to assess synthetic data quality.

\section*{Acknowledgment}
Claude Code (Claude Sonnet 5) was used to help with development and implementation of the Omnisynth codebase. No generative AI models were used for the scientific design, methodology, analysis, or paper writing.

\bibliographystyle{IEEEtran}
\bibliography{references}

\end{document}